\documentclass[aps,twocolumn,prl,longbilbliography,superscriptaddress,shopacs]{revtex4-2} 
\usepackage{graphicx}
\usepackage{bm}
\usepackage{gensymb}
\usepackage[none]{hyphenat}
\usepackage{epstopdf}
\usepackage{multirow}
\usepackage{float}
\usepackage{tabularx}
\usepackage{gensymb}
\usepackage{color}
\usepackage{siunitx}
\usepackage{xr}
\usepackage{amsmath}
\usepackage[colorlinks=true]{hyperref} 
\usepackage{placeins}
\usepackage{multirow}

\newcommand{\sw}[1]{\textcolor{black}{#1}}

\begin{document}
\hspace{5.2in} \mbox{}

\title{Discovery of Charge Order in the Transition Metal Dichalcogenide Fe$_{x}$NbS$_2$}

\author{Shan Wu}
\thanks{These authors contributed equally to this work.}
\affiliation{Department of Physics, University of California Berkeley, California, 94720, USA}
\affiliation{Material Sciences Division, Lawrence Berkeley National Lab, Berkeley, California, 94720, USA}
\affiliation{Department of Physics, Santa Clara University, Santa Clara, CA, 95053}

\author{$\hspace{-0.5em}^{,\dagger\hspace{0.5em}}$Rourav Basak}
\thanks{These authors contributed equally to this work.}
\affiliation{Department of Physics, University of California San Diego, California, 92093, USA}

\author{Wenxin Li}
\affiliation{Department of Applied Physics, Yale University, New Haven, Connecticut 06511, USA}

\author{Jong-Woo Kim}
\affiliation{Advanced Photon Source, Argonne National Laboratories, Lemont, IL, USA}

\author{Philip J. Ryan}
\affiliation{Advanced Photon Source, Argonne National Laboratories, Lemont, IL, USA}

\author{Donghui Lu}
\affiliation{Stanford Synchrotron Radiation Lightsource, SLAC National Accelerator Laboratory   , Menlo Park, California 94025, USA}

\author{Makoto Hashimoto}
\affiliation{Stanford Synchrotron Radiation Lightsource, SLAC National Accelerator Laboratory, Menlo Park, California 94025, USA}

\author{Christie Nelson}
\affiliation{National Synchrotron Light Source II, Brookhaven National Laboratory, Upton, New York 11973, USA}

\author{Raul Acevedo-Esteves}
\affiliation{National Synchrotron Light Source II, Brookhaven National Laboratory, Upton, New York 11973, USA}

\author{Shannon C. Haley}
\affiliation{Department of Physics, University of California Berkeley, California, 94720, USA}

\author{James G. Analytis}
\affiliation{Department of Physics, University of California Berkeley, California, 94720, USA}
\affiliation{CIFAR Quantum Materials, CIFAR, Toronto, Ontario M5G 1M1, Canada}

\author{Yu He}
\affiliation{Department of Applied Physics, Yale University, New Haven, Connecticut 06511, USA}

\author{Alex Frano}
\affiliation{Department of Physics, University of California San Diego, California, 92093, USA}

\author{Robert J. Birgeneau}
\email{Corresponding authors: shanwu@berkeley.edu (S.W.), robertjb@berkeley.edu (R.J.B.)}
\affiliation{Department of Physics, University of California Berkeley, California, 94720, USA}
\affiliation{Material Sciences Division, Lawrence Berkeley National Lab, Berkeley, California, 94720, USA}

\date{\today}
\begin{abstract}
The Fe intercalated transition metal dichalcogenide (TMD), Fe$_{1/3}$NbS$_2$, exhibits remarkable resistance switching properties and highly tunable spin ordering phases due to magnetic defects.  We conduct synchrotron X-ray scattering measurements on both under-intercalated ($x$ = 0.32) and over-intercalated ($x$ = 0.35) samples. We discover a new charge order phase in the over-intercalated sample, where the excess Fe atoms lead to a zigzag antiferromagnetic order. The agreement between the charge and magnetic ordering temperatures, as well as their intensity relationship, suggests a strong magnetoelastic coupling as the mechanism for the charge ordering. Our results reveal the first example of a charge order phase among the intercalated TMD family and demonstrate the ability to stabilize charge modulation by introducing electronic correlations, where the charge order is absent in bulk 2H-NbS$_2$ compared to other pristine TMDs. 
\end{abstract}
\maketitle

\begin{figure*}
\includegraphics[width=2\columnwidth,clip,angle =0]{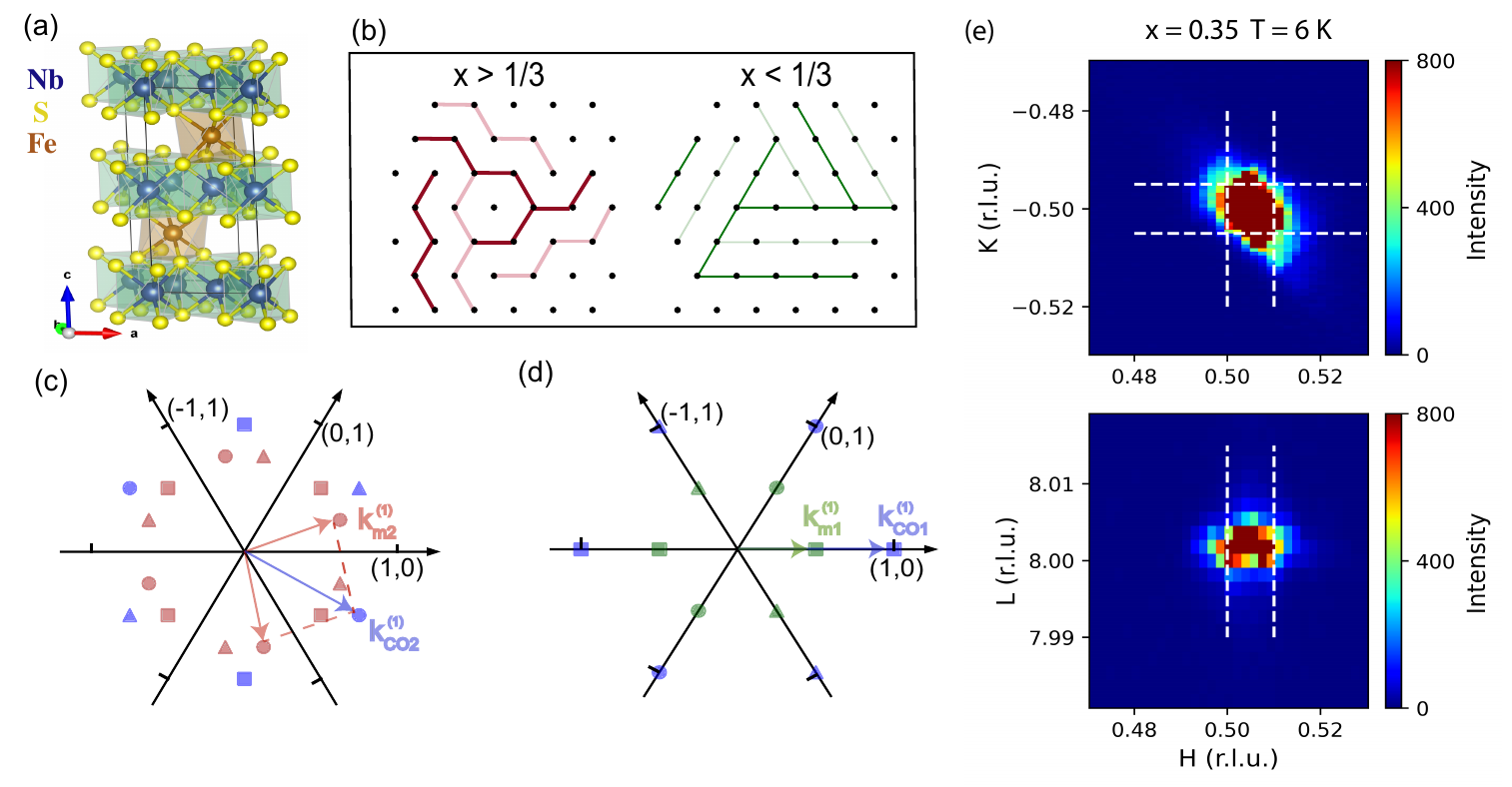}
\caption{\label{fig1} (a) The crystallographic structure of Fe$_{1/3}$NbS$_2$. (b)The stripe ($x < 1/3$, green) and zigzag ($x > 1/3$, red) order spin structures within a single Fe triangular lattice layer, where dark and light lines denote spins up and down. Magnetic wave vectors associated with three magnetic domains in the $(hk0)$ reciprocal lattice plane for the (c) zigzag ($x=$0.35, red symbols) and (d) stripe ordered phases ($x=$0.32, green symbols).  The magnetic wave vector is, for example, $k_{m2}^{(1)}$ = (0.5, 0.25, 0) and $k_{m1}^{(1)}$ = (0.5, 0, 0) in zigzag and stripe phases, respectively. Blue symbols refer to the derived charge order (CO) wave vectors due to the magneto-elastic coupling, \sw{leading to the commensurate CO wave vector value of $\bf k_{CO}$ = (0.5, 0.5, 0) in $x$ = 0.35 sample that agrees with our observation.} The wave vectors associated with three magnetic and charge domains are denoted by circle, triangular and square symbols. (e) The two-dimensional image data in the $(H,K,8)$ and $(H,-0.5,L)$ planes at $T$ = 6 K with the photon energy $E$ = 10 keV reveals a new charge ordering peak at $Q_{CO} =$ (0.5,-0.5,8). }
\end{figure*}

Charge order is an ubiquitous electronic phase in solid-state materials that disrupts translational symmetry by creating a periodic modulation of the charge density \cite{wilson1974,alonso1999,Jerome2002,tranquada1994,Tranquada1995,v.Zimmermann1998,vandenBrink2008,Senn2012,Teng2022}. Exploring its cooperative or competitive interactions with other electronic phases including magnetic order, nematic order, and unconventional superconductivity especially near a putative quantum critical point, has historically generated significant interest in the strongly correlated quantum materials. \cite{kastner1998,kivelson2003,Johnston2010,Fernandes2014,Keimer2015,fradkin2015,Frano2020,Bohmer2022,Fernandes2022}.  

In transition metal dichalcogenide (TMD) materials TA$_2$ (T = Ta, Nb, Mo; A = Se, S), the interplay between electron-electron and strong electron-phonon interactions makes them prototypical examples of charge density wave (CDW) systems and phonon-mediated superconductors \cite{wilson1974,naito1981,castro2001,guillam2008,Rossnagel_2011,Diego2021}. Among the metallic bulk TMDs, NbS$_2$ stands out because none of its polytypes, or those under pressure, have been reported to have a CDW \cite{naito1982,guillam2008,tissen2013,Fisher1980}, in contrast to its isoelectronic and isostructural analogues such as NbSe$_2$ \cite{anjan2013}. The CDW order in bulk 2H-NbS$_2$ is likely quenched by the large anharmonicity of the lattice, resulting in only a faint 1T-polytype-like CDW feature due to a stacking fault \cite{leroux2018,Bianco2019}. Theoretical work suggests that, compared to other TMDs, 2H-NbS$_2$ exhibits stronger many-body effects involving competing Coulomb and electron-phonon interactions and is close to the brink of an instability to charge ordering \cite{Nishio1994,guller2016,heil2017,heil2018,vanLoon2018,Lin2020}.

\begin{figure}
\includegraphics[width=1.\columnwidth,clip,angle =0]{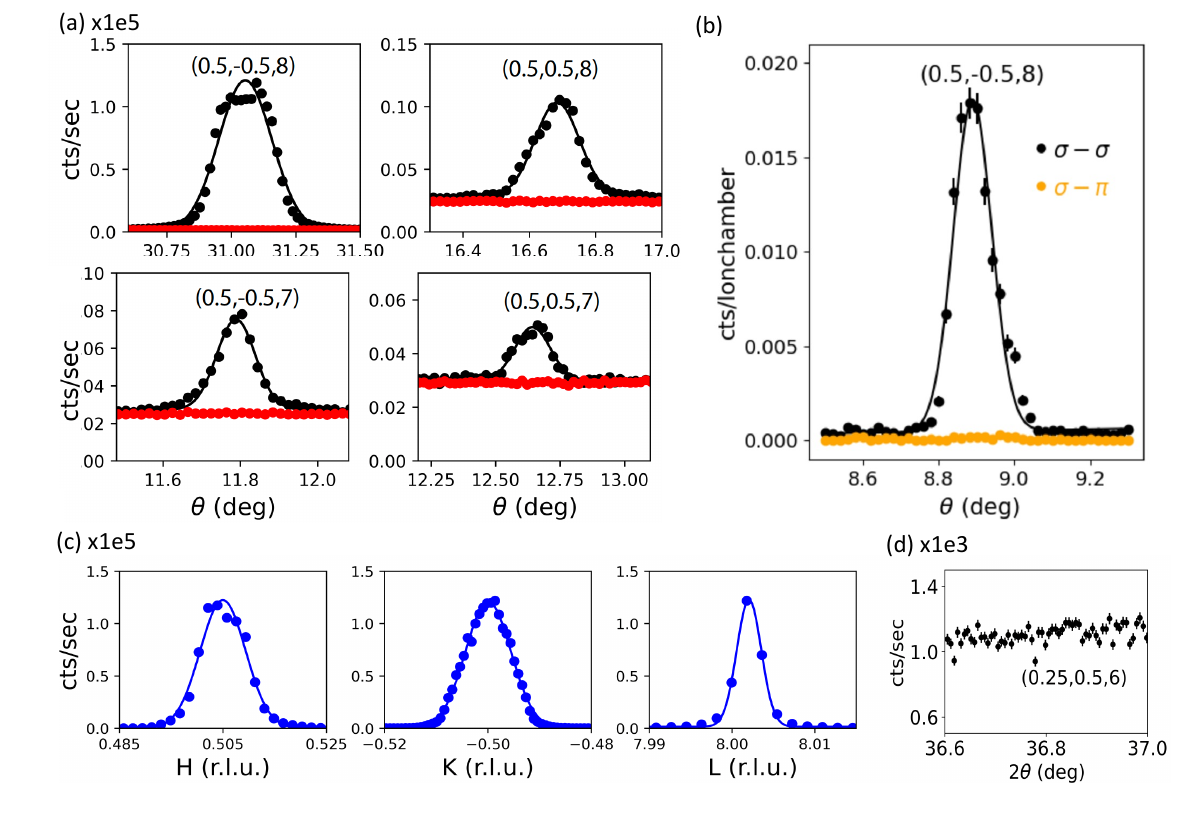}
\caption{\label{COpeak} (a) Representative transverse ($\theta$) scans of charge order peaks measured on one single crystal with $x$ = 0.35 at $T$ = 6 K (black) and 45 K (red). Data were collected on the beamline 6-ID-B at the advanced photon source with incident photon energy $E = 10$ keV. Dots are raw data and solid lines are results of the fits to the Gaussian line shape. (b) \sw{The peak intensity of charge order in the $\sigma-\sigma$ (black) and $\sigma-\pi$ (orange) channel at $T = $ 6 K, bolstering the non-magnetic origin of this peak. The data was collected at the 4-ID beamline at the NSLS-II with the incident energy of 18.95 keV.}  (c) $H,K,L$ cuts across the charge order peak position $\bf{Q_{CO}}$ = ($\frac{1}{2},-\frac{1}{2},8$) from the integration range between the white dashed lines in Fig. 1(e). (d) Longitudinal scan of magnetic peak position at (0.25,0.5,6), related to $\bf{Q_{MO}}$ = (0.25,0.5,0) at $T$ = 6 K. 
}
\end{figure}

TMD are weakly electron-correlated non-magnetic materials.  The weak van der Waals bonding between chalcogen atoms of adjacent layers allows the ready intercalation of magnetic atoms as M$_x$TA$_2$ (M = 3d transition metal) \cite{parkin1980,friend1977}, increasing the electronic correlations. When x = $1/3$, the intercalated atoms typically arrange into a stacked $\sqrt3 \times \sqrt3$ superlattice \cite{boswell1978}, forming a non-centrosymmetric space group $P6_3 22$ with a bi-layer triangular arrangement of M atoms (Fig. \ref{fig1} (a)). This family exhibits a wide range of fascinating magnetic and electronic properties \cite{togawa2012,braam2015,Kousaka_2016,kousaka2009,karna2019, togawa2015,ghimire2018,aoki2019,Park2022}. However, no charge ordering including any CDW phase has been reported in any of the intercalated species so far.

Fe$_{1/3}$NbS$_2$ has demonstrated intriguing current-induced resistance switching features below the Ne$\acute{e}$l transition temperature $T_N \sim$ 45 K \cite{nair2020,maniv2021_1}. The switching behavior which depends on sensitively on the Fe ratio $x$ was later correlated with a highly tunable magnetic phase as the Fe site occupancy changes from vacancies to interstitials \cite{sw_prx2022}. Specifically, under- ($x <$ 1/3) and over-intercalated ($x >$ 1/3) samples reveal distinct antiferromagnetic stripe and zigzag orders, described by the wave vectors $k_{m1}$ = (0.5, 0, 0) and $k_{m2}$ = (0.25, 0.5, 0), respectively (see Fig. \ref{fig1}(b)). Transport measurements revealed a sudden resistivity kink occurring at $T_N$ \cite{friend1977}, and optical polarimetry measurements uncovered three-state nematicity \cite{little2020}. All of these results exhibit the strong spin-charge coupling in this system arising from the duality of localized and itinerant electrons associated with Fe 3$d$ electrons. This naturally raises the question of whether Fe intercalation could stabilize a charge instability and drive charge ordering with enhanced electronic correlations. 

In this paper, we report synchrotron X-ray scattering measurements on Fe$_x$NbS$_2$ samples with under-intercalation ($x=0.32$, magnetic stripe phase with $T_N=34$ K) and over-intercalation ($x=0.35$, magnetic zigzag phase with $T_N=38$ K). We present our discovery of a charge order phase in the over-intercalated sample below $T_N$, which was not observed by previous neutron scattering experiments. We attribute this charge modulation to the coupling between the magnetism and the lattice, and demonstrate that a Fermi surface nesting scenario is unlikely through our angle-resolved photoemission spectroscopy measurements. Our study uncovers the first observation of a charge order phase in the 2H-NbS$_2$ based system and more generally in the intercalated TMD family, \sw{which signifies the disparity from the incommensurate CDW order normally found in the pristine TMDs. }

We used high-quality single crystals with Fe ratios of $x$= 0.32 and 0.35, previously measured with neutron scattering \cite{sw_prx2022}. The X-ray scattering experiments from the temperature of 6 K to 50 K were performed on the 6-ID-B beamline at the Advanced Photon Source (APS) at Argonne National Laboratory. \sw{Additional X-ray scattering measurement with the polarization analyzer was performed on the 4-ID beam-line at the NSLS-II at Brookhaven national lab. } The incident energy used was $E =$ 10 keV. Angle resolved photoemission spectroscopy (ARPES) experiments at base temperature of 10 K were performed at the Stanford Synchrotron Radiation Lightsource beamline 5-2. The measurements were taken between 10 K and 60 K using a photon energy of 160 eV. \sw{We utilized the structure with the space group $P6_322$ \cite{sw_prx2022} as reported before.} For more experimental details see the supplementary information (SI).

We first studied the X-ray scattering on the $x=$ 0.35 sample, which with neutron diffraction showed a zigzag order with a magnetic wave vector $k_{m2}$ = (0.25, 0.5, 0) below $T_N$ = 38 K. At $T = $6 K, we observed new superlattice peaks at $Q$ = (0.5, 0, L) and (0.5, 0.5, L) (L = integer values), as shown in Fig. \ref{COpeak} \sw{(a) \& S2}. We measured six equivalent momentum positions for each $Q$ and confirmed a series of superlattice peaks with half-integer in-plane indices and integer $L$.  At $T=$ 45 K, all those peaks disappear, revealing a temperature dependent behavior.

To address the possibility of a magnetic origin for these peaks, we provide the following observations. We investigated the same $x=$ 0.35 crystal as was used in the neutron scattering measurements: the neutron experiments only revealed magnetic peaks associated with $k_{m2}$ = (0.25, 0.5, 0) and not with $k_{m1}$ = (0.5, 0, 0). Furthermore, these peaks with half-integer in-plane indices are visible at both even and odd values of $L$, contrary to the selection rule of odd $L$ values for the zigzag and stripe magnetic phases, where two adjacent Fe layers stack antiparallel. Finally, the overall temperature-dependent behavior of these peaks differs from that of the magnetic peaks, as discussed below. 

\sw{Moreover, we utilized the analyzer to select different components of X-ray polarization for scattering measurements at both Nb K edge and Fe K edge on the 4-ID beamline at the NSLS-II. With the incident beam featuring $\sigma$ linear polarization perpendicular to the scattering plane, the peak at (0.5, -0.5, 8) exhibits intensity exclusively with the out-going beam featuring $\sigma$ polarization; this is denoted as the $\sigma-\sigma$ channel. The corresponding peak intensity is conspicuously absent within the $\pi$ polarized beam parallel to the scattering plane. The absence of peak intensity in the $\sigma-\pi$ channel excludes the magnetic origin for the superlattice peaks observed in the $x=$ 0.35 sample.} 

We estimated the magnetic intensity to be about 50\% of the atomic Bragg peak intensity at (1, 0, 7) from the neutron scattering measurement. The half-integer peaks in our X-ray measurements, on the other hand, were found to be approximately three orders of magnitude smaller than the atomic peak. These peaks at (0.5, 0.5, $L$) would produce a negligible relative intensity above the background level in neutron measurements. Whereas when interacting with X-rays, the scattering due to charge is inherently stronger than magnetic scattering. In addition, we did not observe any change of the atomic Bragg peaks across the transition (see Fig. \ref{COpeak} (d)). Therefore, we conclude that these superlattice peaks with half-integer in-plane indices and integer L originate from the in-plane staggered charge density.

We investigated the electronic correlation of the charge order peaks by analyzing two-dimensional image data in the $(H,K,8)$ and $(H,-0.5,L)$ planes at $T=6$ K (Fig. \ref{fig1}(e)). By integrating over the intensities marked between two parallel dashed lines, we plotted the $Q$-cuts along the $H$, $K$, and $L$ directions (Fig. \ref{COpeak} (c)). Fitting the peaks to a Gaussian line shape, we extracted a peak width that was comparable to that of the nearby nuclear peak. This demonstrates the presence of three-dimensional long-range charge order, with a estimated correlation length $\xi > $  500 \AA.

\begin{figure}
\includegraphics[width=1.\columnwidth,clip,angle =0]{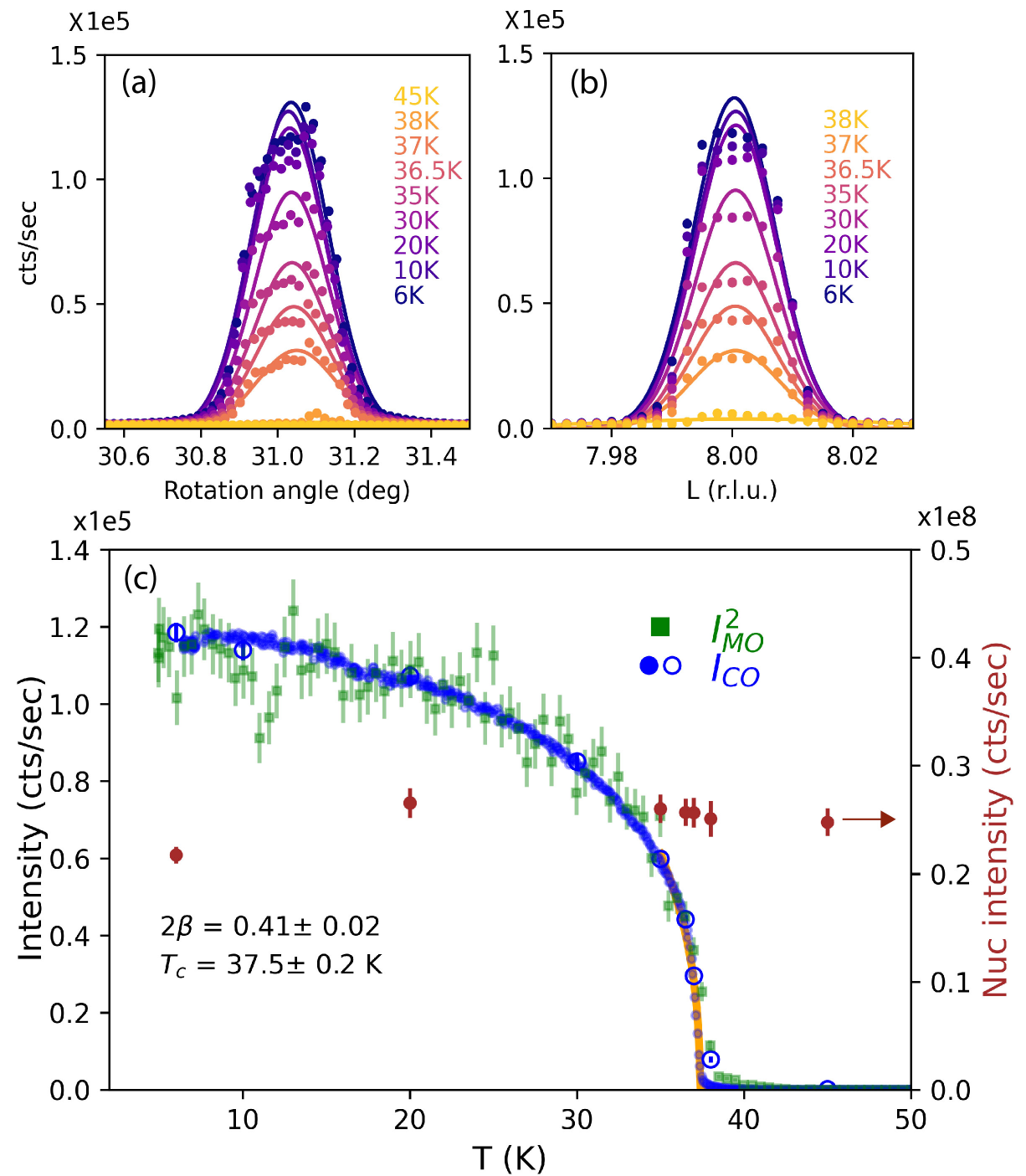}
\caption{\label{OP} (a) Sample rotation angle ($\theta$) scan and (b) $L$ scan of the charge order peak at $\bf{Q_{CO}}$ = ($\frac{1}{2},-\frac{1}{2},8$) in the $x=$ 0.35 sample as a function of temperature.   Solid lines are fits to the Gaussian function. (c) Temperature dependence behaviors for the charge order peak at $\bf{Q_{CO}}$ (blue dots), the atomic Bragg peak at $\bf{Q_{N}}$ = (1,0,7) (brown dots), and magnetic peak $\bf{Q_{MO}}$ = (0.25,0.5,0) (green dots) from previous neutron scattering experiment \cite{sw_prx2022}. The values of $I_{MO}^2$ were scaled by a constant for comparison with the trend of charge ordering intensity $I_{C}$. The empty circles  are integrated intensities from the Gaussian fits in panel (a). The orange line is the result of fit to the power law function of $(1-\frac{T}{T_{C}})^{2\beta}$ for the charge ordering in the temperature range between 30 and 40 K. }
\end{figure}

To study the temperature evolution of the charge order phase, we measured the intensities at the peak position $\bf{Q_{CO}}$ = (1/2, -1/2, 8) as a function of temperature, as shown in Fig. \ref{OP}. Additionally, we performed sample rotation ($\theta$) and $L$ scans at increasing temperatures, which clearly showed the disappearance of the charge order peak at high temperatures. In comparison, no observable change in the atomic Bragg peak at $\bf Q_N$ = (1,0,7) was observed across the transition, consistent with the absence of any structural phase transition.

We determined the transition temperature for the charge order phase by fitting the order-parameter curve with a power law exponent in the temperature range above 30 K. The fit yielded a charge order transition temperature of $T_{C}$ = 37.5(2) K and a power law exponent for the charge ordering of $2\beta$ = 0.41(2). From the magnetoelastic model to be discussed below,  this value implies a $2\beta$ value of 0.21 for the magnetic ordering, consistent with the 2D Ising model value of $2\beta$ = 0.25 \cite{pelissetto2002,onsager}.  The value of $T_{C}$ is in good agreement with the magnetic order transition temperature, suggesting a strong connection between charge and magnetic order.

In contrast, we did not observe any additional temperature-dependent superlattice peaks in the $x = 0.32$ sample that hosts the magnetic stripe phase at $T$ = 6 K. We scanned over roughly 36 peaks with half-integer in-plane indices, as well as other high-symmetry directions, including ($n/4, m/4, L), (n/3, m/3, L), (n/4, 0, L)$ and $(n/3, 0, L)$ ($n,m$ = integers, $L$ = 6,7). We studied two $x=$ 0.32 samples: one neutron sample with a comparable mosaicity to the $x=$ 0.35 sample (FWHM = 0.15$^\circ$), and the second one with a sharper peak width (FWHM = 0.02$^\circ$). Both samples did not reveal any apparent temperature-dependent peak features. The superlattice peak at (0.5,0.5,$L$) persists up to room temperature (see Fig. S1 (b)), most likely due to a small portion of the Fe occupancy at 2$b$ sites in the under-intercalated samples \cite{sw_prx2022}. Given the surveyed momentum positions,  the invisibility of charge ordering in $x=$ 0.32 is plausibly related to the highly tunable magnetic structures with respect to Fe ratio $x$, which will be discussed later.

\begin{figure}
\includegraphics[width=1.\columnwidth,clip,angle =0]{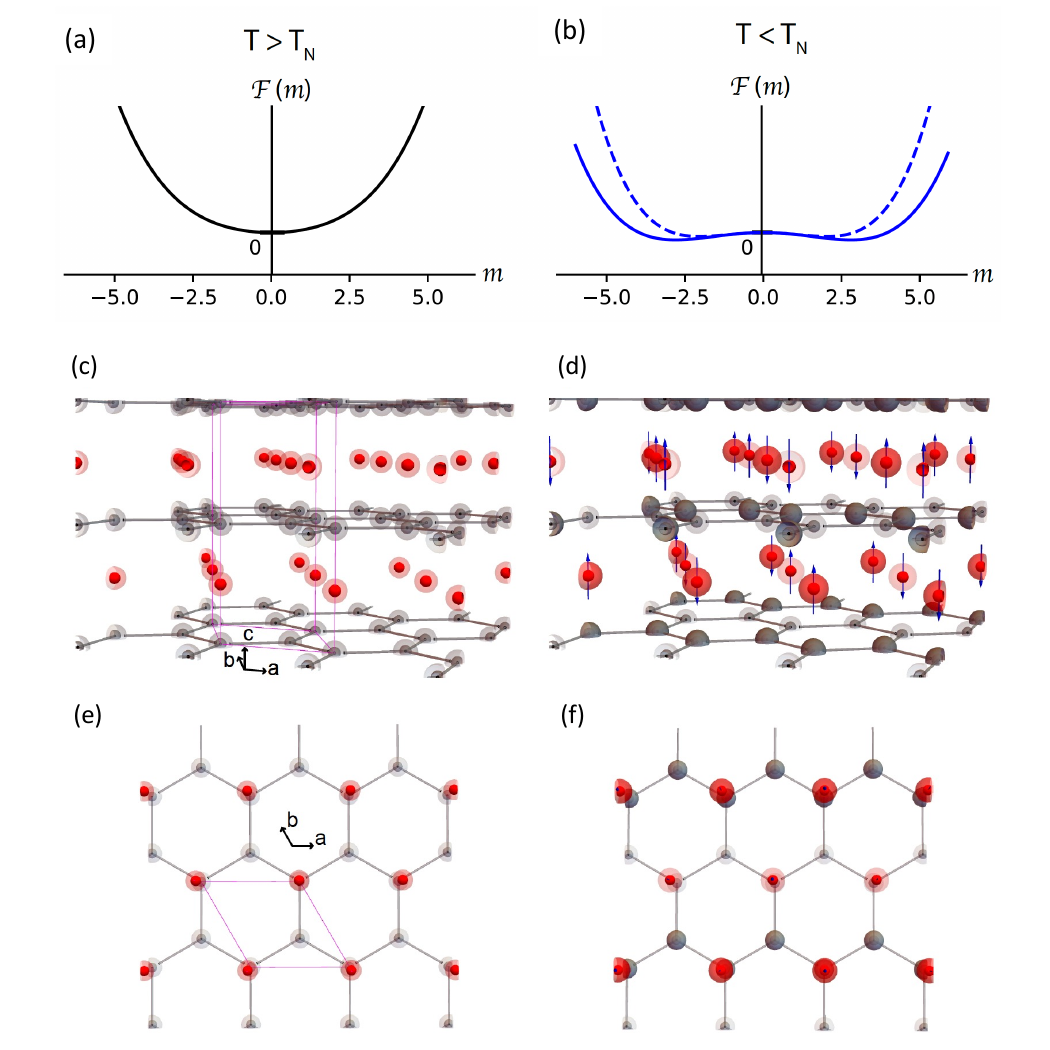}
\caption{\label{fig4} Schematic illustrations of the free energy for (a) $T > T_N$ and (b) $T < T_N$. In panel (b), the dashed blue line is the free energy for the antiferromagnetic order \cite{Stanley1971}. The solid blue line includes the change of the free energy $\Delta\mathcal{F}$ due to magnetoelastic coupling as written in the Eq. (2), which lowering the ground state energy. (c) \& (e) High temperature ($T > T_N$) lattice showing only Fe (red) and Nb (dark) atoms \sw{ in three-dimensional and $ab$-plane projections, respectively. The pink line denotes the unit cell.} (d) \& (f) Illustration of the 3D charge ordering ($T < T_N$), accompanied by the zig-zag spin ordering \sw{in three-dimensional and $ab$-plane projections, respectively.} The dark and light red (grey) spherical clouds denote the charge density modulation localized on Fe and Nb sites, and blue arrows mark the spin structure of Fe. \sw{This charge order is associated with the wave vector $k_{CO} = (1,-0.5,0)$ induced by the corresponding magnetic wave vector $k_{M} = (0.5,0.25,0)$.}   }  
\end{figure}

The charge order in the zig-zag phase is clearly driven by the antiferromagnetic order. To further understand the intimate relation between the charge and magnetic orders, we compared the temperature-dependent intensities of the two orders (Fig. \ref{OP} (c)). Interestingly, the square of the magnetic intensity $I_{MO}^2$ matches perfectly with the charge order intensity $I_{CO}$ after scaling by a constant factor. This correlation can be attributed to the charge modulation from magnetoelastic coupling between the lattice and magnetism \cite{callen1963}, which induces uniform strain $\epsilon$ via the coupling term:
\begin{equation}
\sum \epsilon_{i,j} m_im_j.
\end{equation}
The free energy for the magnetic order is in the form of $\mathcal{F}(m) = -a_m(1-\frac{T}{T_N})m^2 + b_mm^4 (a_m, b_m > 0)$ \cite{Stanley1971}. Because of the magnetoelastic coupling (Eq. 1), there is an additional term $\Delta \mathcal{F}$ in the free energy \cite{Harris1997}, 
\begin{equation}
\Delta\mathcal{F} \sim -\gamma \rho m^2 + \frac{1}{2} \rho^2,
\end{equation}
where $m$ is the magnetic moment, $\gamma$ is the coupling strength and $\rho$ is the modulated charge density, which is a secondary order parameter. This coupling leads to the charge scattering intensity from minimizing the free energy (Fig. \ref{fig4} (b)) as:
\begin{equation}
I_{CO} \sim \rho^2 \sim m^4 \sim I_{MO}^2. 
\end{equation}
This agrees with the empirical intensity relationship in our data.

In $Q$-space, the free energy can be expressed as:
\begin{equation}
\sum_{k_1,k_2} \rho(-k_1-k_2)m(k_1)m(k_2) + \frac{1}{2}\rho^2
\end{equation}
This formula implies that the wave vector of the charge order, $k_{CO}$, would be the sum of two magnetic wave vectors, $k_m$. As shown in Fig. \ref{fig1} (c) in the $x=0.35$ sample, 12 equivalent magnetic wave vectors are assigned to three magnetic domains (denoted by three symbols). Within one domain, two $k_{m2}$ (red vector) can generate one charge wave vector $k_{CO2}$ (blue vector), leading to a total of six equivalent wave vectors associated with (0.5, 0.5, 0). This naturally explains the observed peak positions with half-integer in-plane indices considering the momentum positions $Q_{CO2} = \tau \pm k_{CO2}$ ($\tau$ is the atomic Bragg position). For the stripe phase, the magnetic wave vector occurs at half-integer values (Fig. \ref{fig1} (d)). Any charge scattering originating from the magneto-elastic coupling would be orders of magnitude weaker than the regular Bragg scattering from the crystal lattice. Doubling of $k_{m1}$ causes the charge order peaks to appear at the atomic Bragg peak positions, making them undetectable in the experiment. Consequently, we are unable to determine if any magnetoelastically induced charge ordering in the $x$ = 0.32 stripe phase sample. 

The magnetic order induced charge ordering in Fe$_{0.35}$NbS$_2$ is different from that of other pristine TMDs \sw{exhibiting incommensurate charge-density-wave order}, such as 2H-NbSe$_2$. \sw{2H-NbSe$_2$ was reported to display the low-energy Fermi surface nesting \cite{Kiss2007,borisenko2009,rahn2012,straub1999,rossnagel2001,valla2004,shen2008}; but later the CDW formation was found to be attributed to the electron-phonon coupling mechanism \cite{weberprl2011,weberprb2018}.}
In Fe$_{0.35}$NbS$_2$, our ARPES measurements did not reveal any clear evidence of CDW features via auto-correlated Fermi surface analysis \cite{hashimoto2011reaffirming}  (see Fig. S3). In addition, our spectra have revealed a strong photon energy dependence of the electronic structure caused by the Fe intercalation, distinct from the quasi-2D electronic band in bulk NbS$_2$ \cite{youbi2021}. These results confirm that the charge order here is not a simple low-energy electronically driven transition, but rather the intra/interlayer magnetic exchange coupling involving Fe is a vital factor.

To conclude, we have conducted synchrotron X-ray scattering experiments on Fe-intercalated 2H-NbS$_2$ transition metal dichalcogenide, Fe$_x$NbS$_2$, which has been recently reported to exhibit highly tunable magnetic phases across $x =$ 1/3. Our study has revealed a new three-dimensional charge order phase with a wave vector of $\bf k_{CO}$ = (0.5, 0.5, 0) in the $x =$ 0.35 sample ($>$ 1/3, Fe interstitial), accompanied with the antiferromagnetic ordering. The 3D charge ordering pattern is illustrated as in Fig. \ref{fig4} (d). The temperature dependence of the intensity suggests a strong magneto-elastic coupling effect as the origin of the charge order. In contrast, we did not observe superlattices of charge ordering in the $x =$ 0.32 sample ($<$ 1/3, Fe vacancies), which would likely be \sw{obscured} beneath the \sw{strong} atomic Bragg peaks if they were magnetically driven. \sw{Recently, the magneto-elastic coupling is suggested to amplify the non-local resistance switching feature in this system \cite{Haley2023}, which culminates the potential importance of the intertwined charge and magnetic orders in influencing the switching effect.} The microscopic mechanism of this charge ordering, the role of magnetic defects, and the connection with intriguing resistance switching effects call for further experimental and theoretical investigation. Our work opens up more opportunities to explore spin and charge ordering coupled phenomena in correlated triangular lattice antiferromagnet. 

This work is funded by the U.S. Department of Energy, Office of Science, Office of Basic Energy Sciences, Materials Sciences and Engineering Division under Contract No. DE-AC02-05-CH11231 within the Quantum Materials Program (KC2202). This material is based upon work supported by the National Science Foundation under Grant No. DMR-2145080. A.F. was supported by the Research Corporation for Science Advancement via the Cottrell Scholar Award (27551) and the CIFAR Azrieli Global Scholars program. The photo-emission work is partially supported by National Science Foundation under Grant No. DMR-2239171. This research used resources of the Advanced Photon Source, a U.S. Department of Energy (DOE) Office of Science user facility operated for the DOE Office of Science by Argonne National Laboratory under Contract No. DE-AC02-06CH11357. Use of the Stanford Synchrotron Radiation Lightsource, SLAC National Accelerator Laboratory, is supported by the U.S. Department of Energy, Office of Science, Office of Basic Energy Sciences under Contract No. DE-AC02-76SF00515. This research used beamline 4-ID of the National Synchrotron Light Source II, a U.S. Department of Energy (DOE) Office of Science User Facility operated for the DOE Office of Science by Brookhaven National Laboratory under Contract No. DE-SC0012704.

\bibliography{ChargeorderFexNbS2.bbl}
\end{document}